# Impact of Exoplanet Science on Society: Professional Contributions, Citizen Science Engagement and Public Perception

Hans-Jörg Deeg
Instituto de Astrofísica de Canarias, C. Via Lactea S/N, La Laguna, Tenerife, Spain
e-mail: hdeeg@iac.es

## Abstract

The impact of exoplanet science on both the scientific community and on the general public is presented through various indicators and examples. It is estimated that about 3-4% of all refereed astronomy articles focus on exoplanets, and between 15-20% percent of current, and up to 25% of upcoming astronomy space missions are dedicated to exoplanet research. Also, about 15-20% of the science cases for large multi-purpose ground-based astronomical instruments involve exoplanet science.

Interactions between the scientific community and the public occur on several levels and play a crucial role in shaping the future of exoplanet science. The rise of citizen science platforms and the successes of coordinated observing projects involving amateur astronomers have engaged the public in meaningful scientific contributions, and contribute to some areas of discovery and characterization of exoplanet systems, for which several examples are given. These initiatives not only fuel public interest in the search for extraterrestrial life but also promote STEM education, broadening participation in science.

Lastly, the changing perception of the informed public about the existence of 'other Earths' and life in the Universe in the light of results from exoplanet science is outlined. Media coverage of results from exoplanet science has furthered the acceptance that extraterrestrial life, be it intelligent of not, is not rare in the Universe. The shift in perception that such life might be detected in a potentially not very distant future has, in turn, promoted public support for the research infrastructure necessary to sustain the growth of exoplanetology.

## Introduction

Since the discovery of the first exoplanets in the 1990s, the field of exoplanet science, or 'exoplanetology,' has undoubtedly become a central component of astronomy and one of its most rapidly advancing areas. But now, well into the 21st century, how far has the impact of exoplanet science reached, both in the professional domain and in the society at large? Which interactions with society has it generated? These questions are significant not only for researchers in the field of exoplanets, who seek to understand the broader impact of their work, but also for policymakers responsible for making high-level decisions regarding support for related research activities.

The first part of this chapter provides several indicators of exoplanet science's current role within professional astronomy. However, assessing the broader impact on the general public is more complex, with media playing a key role in shaping public perceptions. We present public engagement on multiple levels, beginning with the involvement of interested individuals in citizen science projects, followed by an examination of how media coverage influences the general public's perception of 'other worlds' and the possibility of life in the universe.

**Impact of exoplanet science within professional astronomy.**

Since the discovery of the first widely recognized exoplanet in 1995, it is without doubt that exoplanet research has become a significant field in astronomy. But how big has it become in absolute terms, and in relative terms, against the entire field of astronomy and astrophysics? Here we provide a few indicators that may quantify this issue.

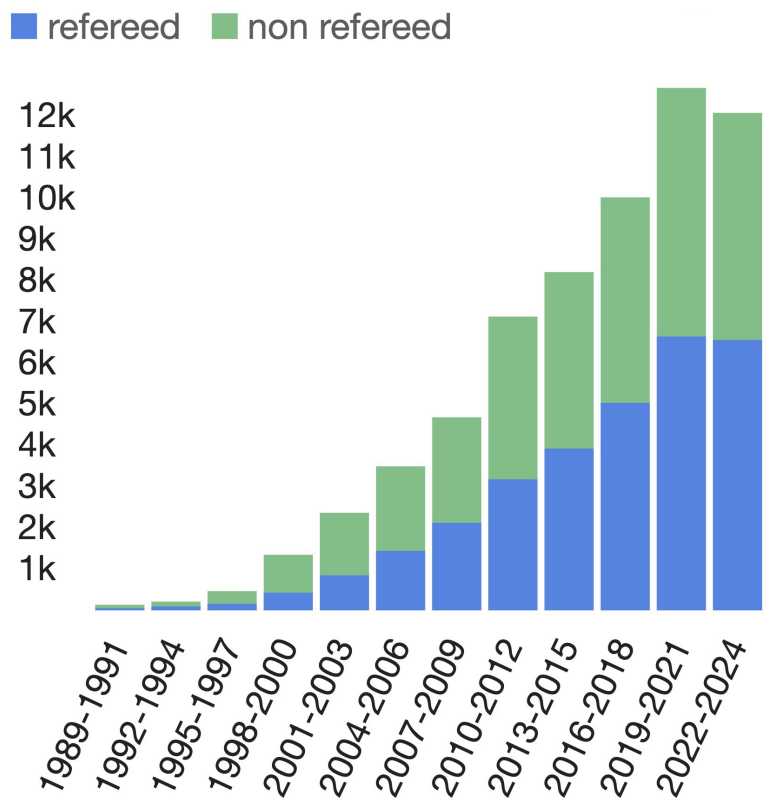

**Fig.1** Number of publications, in 3-year bins from 1989 to 2024, that contain the terms 'exoplanet' or 'extrasolar AND planet' in abstract or main-text. Data extracted from the astronomy collection of the Astrophysical Data System on 2024 Sept. 4.

Fig. **1** shows the number of scientific publications that contain the terms 'exoplanet' or 'extrasolar AND planet', be it in their abstract or their main text, from a search in the astronomy collection of the Astrophysical Data System (ADS). Starting from a rather constant level of 10-15 refereed papers per year during the 1980s and a moderate rise in the 1990's, the field started to take off in 1998 or 1999. This was followed by a constant increase in productivity during the first two decades of the 2000's, surpassing 1000 refereed papers per year in 2011 and 2400 papers in 2022.

To evaluate the relative productivity of exoplanet science within the whole field of astronomy, Fig. **2** shows the yearly fractions of refereed papers found by above search, versus all papers in ADS' astronomy collection. Until 1993, exoplanets were a marginal topic mentioned in less than 0.2% of all papers. This changed after the discovery of the first exoplanets (around pulsars in 1992 and main-sequence stars in 1995). Since the late 1990s, a roughly linear increase has set in, and in the mid 2020s, this fraction is approaching 8%. Exoplanets are therefore truly a science of the 21$^{st}$ century!

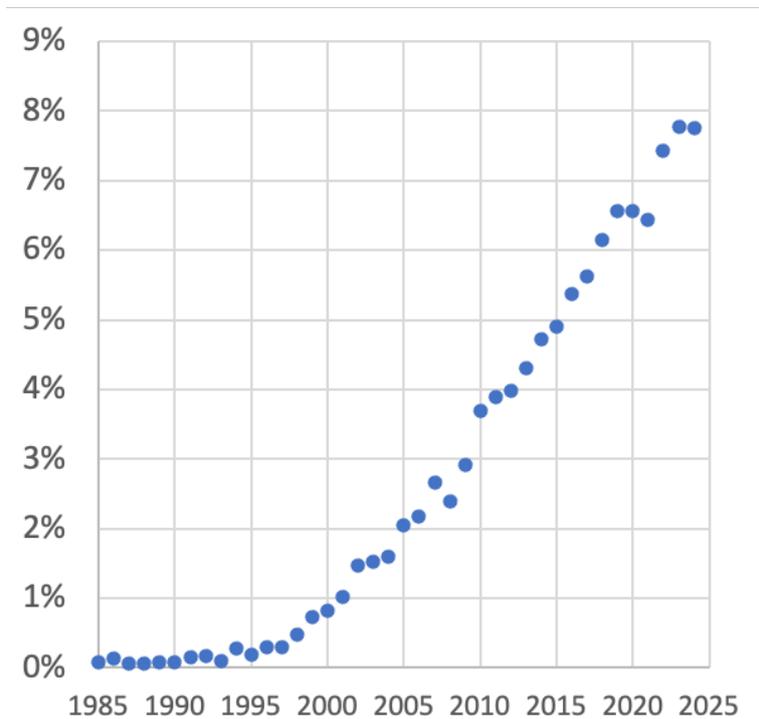

**Fig. 2** Fraction of refereed publications that contain the terms 'exoplanet' or 'extrasolar AND planet' in abstract or main-text, relative to all refereed publications in the astronomy collection of ADS.

Among the exoplanet-papers found in this search, many of them mention exoplanets however only peripherally, and only an estimated 40-50% have a *primary* focus on exoplanets. This fraction was obtained from a manual revision of the titles (and abstract, if needed) of the last 100 papers from 2023 that were returned by the search, from which only 38 were deemed to focus on exoplanets or on closely related analytical or instrumental techniques. An identical exercise with 100 publications from the year 2016 found 55 papers to focus on exoplanets (Deeg & Belmonte 2018). In conclusion, about 3 – 4% of the current technical works in astronomy are focused onto – or driven by – exoplanet science.

A further indicator on the impact of exoplanetology within the entire field of astronomy might be the number of conferences related to exoplanets. A review of the titles of all IAU Symposia of the last 10 years (Symposia 306 to 394, covering May 2013 to August 2024) shows that eight of these symposia (about 9%) treated exoplanets as a main topic. Similarly, an analysis of large meetings covering all of astronomy during the 2010's (IAU General Assemblies, American Astronomical Society meetings, European Astronomical Society meetings) showed that about 10% of such meetings were dedicated to exoplanets (Deeg & Belmonte 2018).

One of the most important consequences of exoplanet science is its influence onto advanced astrophysics instrumentation, both ground and space-based. This is also an indicator of current and future funding that is expended for this science.

Only relatively small ground based facilities are primarily dedicated to exoplanet science (mainly several transit surveys), whereas exoplanets have been a strong driver in the large expenditures needed for space missions. To date (2024), the following missions with a primary dedication to exoplanets have been put into operation (see chapter -> Space Missions for Extrasolar Planets: Overview and Introduction):

- CoRoT (active 2007-2012), led by the French space agency CNES, a dual exoplanet-detection and asteroseismology mission.
- Kepler (2009-2012, relabeled as K2 until 2018) by NASA, the most successful exoplanet finder to date.
- TESS (currently active, launched 2018) by NASA, an all-sky survey to detect exoplanets.
- CHEOPS (currently active, launched 2019) by ESA, primarily dedicated to exoplanet follow-up.

Of significant impact in exoplanet science, but not originally or primarily designed for it, are also the Hubble Space Telescope (launched 1990), the Spitzer space telescope (main phase 2003 - 2009, with limited capabilities until 2020), the Gaia mission (launched 2013 and expected to revolutionize astrometric detections of exoplanets), and of course the James Webb Space Telescope (launched 2021), with two of its four principal science cases being strongly related to exoplanets: the *birth of Stars & Protoplanetary Systems* and *Planets & Origins of Life* (https://jwst.nasa.gov/science.html).

In order to place the number of exoplanet-related space missions into context against all operating astrophysics-related missions, NASA's astrophysics program (NASA 2024) currently lists 14 missions in active status (as operating or extended missions), while ESA maintains 6 operational astrophysics missions (ESA 2024). Depending on the extent to which multi-purpose missions like JWST are considered to be dedicated to exoplanets, we may estimate that 15 - 20 % of currently active astrophysics space missions are dedicated to exoplanets.

Regarding missions in development, ESA indicates 10 missions with this status. Most of them are small missions, but it is of note that its two upcoming mid-sized ('M') missions are both dedicated to exoplanets (PLATO for launch in 2026 and Ariel in 2029); exoplanets represent therefore a large fraction of ESA's current efforts. NASA lists 11 astrophysics missions in the implementation phase, of which the Nancy Grace Roman Space Telescope (launch 2026) is partially dedicated to exoplanets; their only fully exoplanet-dedicated mission is however the ESA-led Ariel (which has a NASA contribution). The Chinese Academy of Science advances the Earth 2.0 or ET transit survey with an expected launch in 2028, which will re-observe the Kepler-field (Ge et al. 2022, 2024); little information is however available about China's wider astrophysics-related space programme. Overall, we may expect that exoplanet science is driving a similar or somewhat larger fraction of upcoming (with secured launch) astrophysics space missions than currently in operation, estimated at 20 - 25 %.

Lastly, exoplanet-related high-level science cases for the largest actual projects for ground-based multi-purpose astronomical instrumentation are indicated:

- Atacama Large Millimeter Array (ALMA): *Planet-forming disks* is given as one of six science themes (https://almascience.nrao.edu/alma-science)
- European Extremely Large Telescope (E-ELT): *Exoplanets — Towards other Earths* is one of its six principal cases (Kissler-Patig & Lyubenova 2011)
- Thirty Meter Telescope (TMT): *The Birth and Early Lives of Stars and Planets* and *Exoplanets* are two out of nine of its science cases (Skidmore 2015)
- Square kilometre array (SKA): *Seeking the origins of life*, with studies on planet and star formation and also of SETI, is one of nine science goals of this radio telescope project (https://www.skao.int/en/explore/science-goals)
- Vera C. Rubin Observatory (formerly LSST): No specific exoplanet science goal is listed (https://rubinobservatory.org/explore/science-goals), but its large-scale surveys may contribute to the discovery of transiting exoplanets (Lund et al. 2015, Tamburo et al. 2023).

From this, a dedication to exoplanet science of about 15% for these major ground facilities can be estimated.

In summary, exoplanet science currently contributes 3-4% of all publications within the broader field of astrophysics, with this share continuing to grow. It also accounts for approximately 15-20% of ongoing ground and space instrumentation projects, and potentially a higher proportion (20-25%) of upcoming space missions. This significant contribution from a relatively young field reflects the perceived potential of exoplanet science, not only within professional astronomy but also among the informed public and policymakers. The focus on the discovery of Earth-like planets and, more broadly, the search for signs of life beyond Earth plays a key role in driving public interest and securing political and financial support.

## Exoplanets and the general public

Assessing the impact of exoplanet science onto the broader population is more challenging due to the lack of quantitative indicators. In this section, we highlight contexts where exoplanet research reaches and engages the informed public—a group with awareness of natural sciences and capable of understanding fundamental scientific concepts.

### *Citizen Science*

Several levels of involvement by non-professional actors may be encompassed as citizen science. Among this, we include also activities oriented towards outreach or within pre-academic educational stages, where astronomy-related activities are frequently offered with the aim make young people interested in wider further STEM-related fields (science, technology, engineering and mathematics) and in the related careers (Yüzgeç & Okuşluk 2023, Onuchukwu et al. 2024).
The involvement by citizen scientists may go from short interactions, like the installation of a screensaver for some distributed computing project, to a very serious level that approaches the work of professionals, such as displayed by some advanced amateur astronomers.

**Distributed computing and analysis projects** are one of the most common involvements of citizens in actual science projects. On the most basic level, also known as volunteer computing, a citizen lends private computing power to a science project employing distributed computing. The project seeks to solve a problem which is difficult or infeasible to tackle using other methods, but it does not require further citizen interaction besides the installation of a software running on a networked computer. A historical example of this was SETI@home, a search for signals from extra-terrestrial intelligences in radio data obtained by SETI projects at the Arecibo and Green Bank radio telescopes. SETI@home was released to the public in 1999 and active until 2020, and its iconic screensaver (Fig. **3)** was a common sight on many computers in the first years of the 21$^{st}$ century. Non-interactive volunteer computing projects may have however declined in appeal, partly because advancements in computational power have reduced the need for them, and also because they offer limited educational benefits.

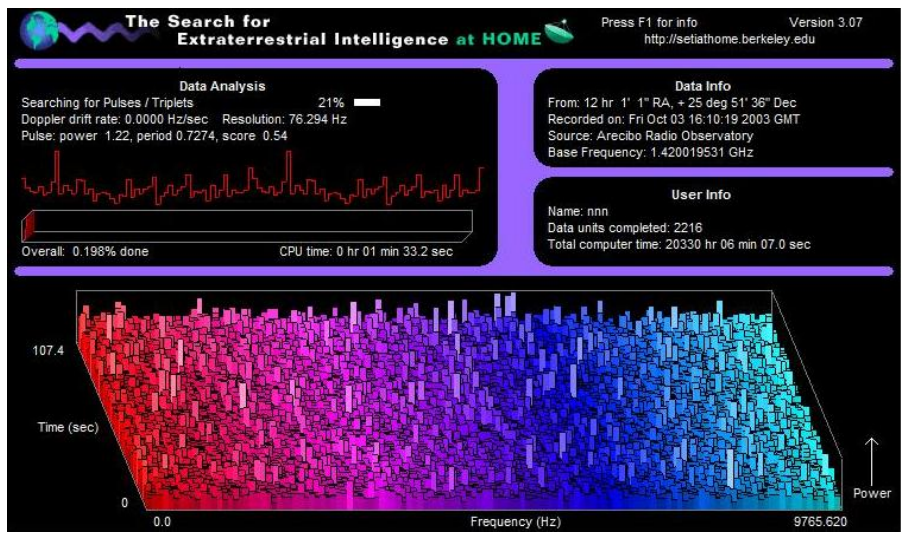

**Fig 3**: The SETi@home Classic screensaver and distributed computing project in its original ('classic') version. (Source Wikimedia, SETI@HOME is licensed under the GNU General Public License)

On the next level of citizen involvement are projects that are computer-based but require active participation. Typically, repetitive tasks that are difficult for computer codes but easy for the human brain are being distributed to the participants. The Zooniverse (www.zooniverse.org; Lintot 2019), which has been a pioneer of this concept, currently (Sept. 2024) hosts 84 projects from a variety of disciplines, mostly involving classification tasks. A fairly large fraction of them, 21, are related to space sciences of with the following being more directly about exoplanet systems: 'Exoasteroids' – a search for asteroids around white dwarf stars, 'Disk detective' – a search and analysis of circumstellar, and potentially protoplanetary disks, and 'Planet Hunters NGTS', which is the latest incarnation of the successful Planet Hunters project. The first version of Planet Hunters was launched in 2010 by a team from Yale University, where participants were employed to spot transit-like features in chunks of light curves from the Kepler space mission. In a second phase, they were also tasked to inspect and confirm the found candidates, with an assessment if further follow-up is warranted (Schwamb et al. 2012). Planet Hunters was later modified to analyze data from the K2 (Kepler's successor) and TESS space missions, while it is currently focused on the ground-based Next Generation Transit Survey (see chapter 'Transit Photometry as an Exoplanet Discovery Method' for an overview over transit detection projects). Planet Hunters has been one of the most successful citizen science project in exoplanetology; it has contributed to the discovery of over 10 exoplanets, of which two are named after it: PH1 b and PH2b, with PH1 b being a relatively rare circumbinary planet (Schwamb et al. 2013).
A further noteworthy project is the Exoplanet Explorer (with uncertain status in 2024, listed in Zooniverse as 'out of data'), which also analyzed data from the K2 mission. With over 15000 volunteers it aided in the discovery of several planet systems, including one with five transiting planets, named K2-138 (Christiansen et al. 2018). The most impacting result of such projects to date might however have been the discovery of a mysterious object with very unusual brightness variations (Fig. **4**) by members of the Planet Hunters project (Boyajian et al. 2016). A variety of explanations have been brought forward (see Wright & Sigurðsson 2016 for an overview), including very speculative ones involving extra-terrestrial intelligences, which led to significant media attention. The discovery of this object, also known as 'Boyajian's star', has been a wonderful result of the power of citizen science: its light curve had been discarded by the algorithms that previously had sifted through Kepler data, and only through the viewing by real persons was its strange nature recognized, leading to one of the strangest puzzles in current-day astronomy.

Further resources on space-related citizen-science projects, we also refer to dedicated web-sites at NASA ( https://exoplanets.nasa.gov/citizen-science/) and ESA ( https://www.esa.int/Enabling_Support/Preparing_for_the_Future/Space_for_Earth/Citizen_science ). Data from upcoming exoplanet-related missions, like PLATO, the Roman Space Telescope or the Habitable Worlds Observatory will also become available to both the scientific community and to the public, and will provide ample opportunities for future citizen science involvement.

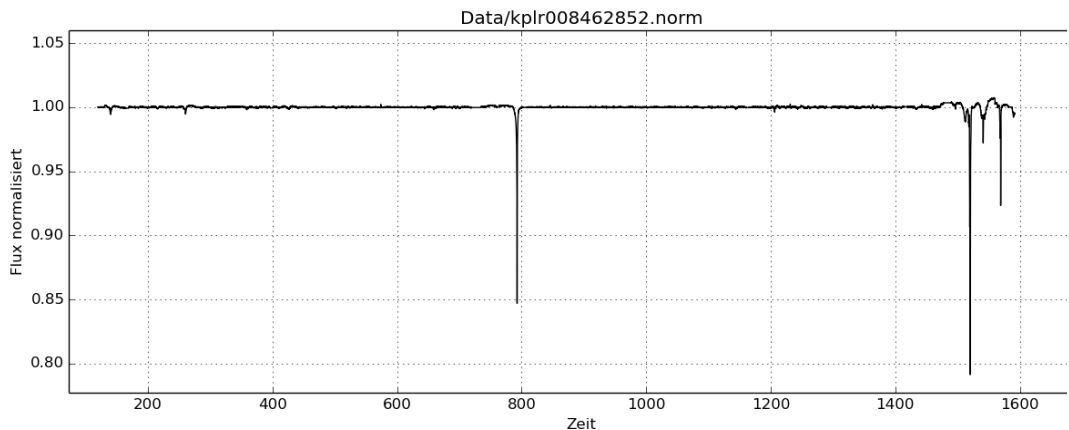

**Fig 4**: The unusual light curve of Boyajian's star (KIC 8462852) which was discovered by Citizen Scientists in data of the Kepler space mission. (Figure by JohnPassos - Own work, CC BY-SA 4.0, https://commons.wikimedia.org/w/index.php?curid=46685223 )

**Amateur astronomers** in their classical form, with their proper observations of the sky, are also contributing significantly to exoplanet science. In contrary to expectations prior to the discovery of the first transiting planets, exoplanet observations do not exclusively require large professional equipment, instead, small instruments accessible to amateurs may be useful as well. Indeed, several pioneering projects to search for transiting planets were based on amateur equipment – although executed by professionals – and one of the two groups that simultaneously detected the first exoplanet transits, of HD209458 b, used a small 10 cm telescope (Charbonneau et al. 2000). This discovery gave immediately rise to many successful transits observations by amateurs, for which the Exoplanet Transits Database (ETD, Poddany et al. 2010, http://var2.astro.cz/ETD) and the database of the American Association of Variable Star observer (AAVSO, https://www.aavso.org/ ) have become the principal repositories. Amateur observations of transits have been aided by the availability of detailed observing instructions (e.g. Gary 2010, Dennis 2016) and by freely available software geared towards differential precision photometry by non-professionals, such as AstroImageJ (https://www.astro.louisville.edu/software/astroimagej, Collins et al. 2017), C-Munipack (Motl 2024) or SIRIL (Team-FreeAstro 2024). Some of the citizen science projects have also developed their specific analysis software, and many light curves from amateurs are good matches to data taken by professional observatories (Fig. **5**).

The main driver for amateur exoplanet observations is the follow-up of known transiters, with the principal objective being the maintenance or refinement of transit ephemeris. Many exoplanets that were discovered from observations made over shorter time spans (several months or less), like those from CoRoT and most from TESS, have transit-ephemeris with a severely limited precision. Within a few years, transit times will then accumulate uncertainties so big (more than 2-3 hours) that further transit observations cannot be planned well. Prior transit re-observations are hence used to generate improved ephemeris, which in turn permit efficient reobservations by future ground and space-based instrumentations over time-scales of years to decades (Dragomir et al. 2020, Deeg et al. 2020, Zellem et al., 2020). The importance of ephemeris maintenance and refinement is epitomized by the automatic ephemeris fitting in ETD, which is based on quality-weighted sets of the uploaded transit observations. Amateur data have been taken into account in several publications presenting improved transit ephemeris; e.g. for planets found by CoRoT (Klagyivik et al. 2021), TESS (Peluso et al. 2023) or WASP (Noguer et al. 2024). Transit re-observations may also give rise to the detection of transit timing variations (TTVs) – these are deviations from strict periodicity that indicate non-linear ephemeris. TTVs may be due to the presence of further orbiting bodies and/or relevant star-planet interactions; they may also be used to estimate the masses of transiting planets (see chapter Transit-Timing and Duration Variations for the Discovery and Characterization of Exoplanets). An example is CoRoT-11 b, which indicates TTVs from both amateur and professional observations (timings in ETD from 2011- 2024 and Deeg et al. 2020). ETD also tracks variations in transit depths and durations, although conclusive evidence for the presence such variations based on amateur data is more difficult to obtain, given that these parameters are very sensitive to data quality.

A selection of observational citizen science projects that pursue these and related topics are currently Exoplanet Watch (https://exoplanets.nasa.gov/exoplanet-watch), the Unistellar Network (https://science.unistellar.com/, Peluso et al. 2023) – a private /public partnership supplying also suitably fitted small telescopes (Marchis et al. 2020) – and Exoclock (https://www.exoclock.space/, Kokori et al. 2022, 2023), with the latter one dedicated to the ephemeris-refinement of target planets for the future ARIEL space mission.

A further driver for amateur involvement is also the identification of false alarms among planet candidates. Such identifications are needed if the photometry from the discovery-survey (be it ground or space-based) is based on imaging of low spatial resolution. In this case, follow-up photometry from higher resolution imaging is needed, in order to distinguish if a planet-candidate displays a transit on the target star, or if a deeper eclipse appears on a nearby star, which means the presence of a false alarm. It is of note that such follow-up observations are also meaningful for candidates with transits that are too shallow to be detected again, as long as the presence of eclipses at all nearby stars can be excluded with certainty. Such work by amateurs was already included in the verification of an exoplanet by the ground-based XO planet survey (McCullough et al. (2007), while more detailed theoretical and observational foundations for these follow-up observations are described by Deeg et al. (2009). For candidates from the TESS mission, the aforementioned Unistellar Network and others are currently pursuing false alarm identifications with the help of citizen observers; see Sgro et al (2024) for a successful example. For the upcoming PLATO transit survey mission (Rauer et al. 2010, 2024 and chapter Space Missions for Exoplanet Science: PLATO), to be launched in 2026, a dedicated program for ground-based photometric follow-up is currently prepared, with branches for professional and for amateur observers, including outreach organizations (Deeg & Alonso 2024, https://citizen.plato-planets.at/ ).

Proposed further objectives of amateur exoplanet observations are also the confirmation of long-periodic planets (by re-observation of their transits), the search for transits of planets that are currently only known from radial velocity detections, or the pursuing of microlensing events caused by exoplanets (Peluso et al. 2023). Boyajian's star, the prominent discovery made from Kepler light curves by citizen scientists, has also become a target of intense surveillance by amateur observers: the database of the AAVSO lists presently (2024 Oct 2) 137 345 observations by 137 observers, with new observations arriving most of the days.
While photometric observations are clearly the main subject of amateur involvement in exoplanetology, some private initiatives for SETI searches in the optical (e.g. Schuetz 2018) and radio domains (Project Bambi, www.bambi.net/) are also of note.

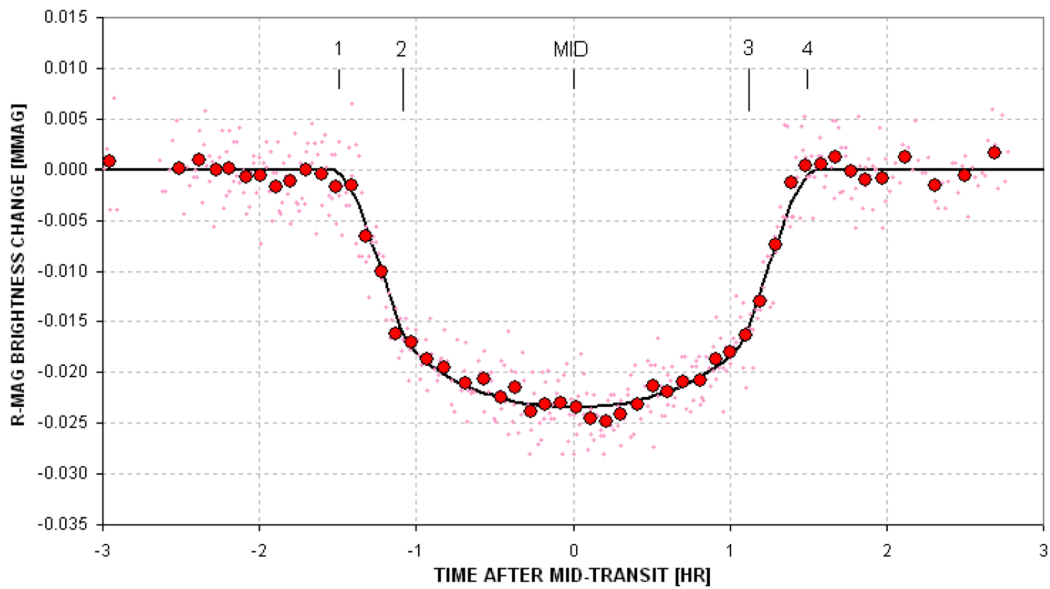

**Fig. 5**. Amateur light curve of a planet transit across the 11.3 mag star XO-1, made with a 14-inch telescope. From Gary (2010), reproduced with permission by the author.

*The Public's View About the Presence of Life in the Universe*

The general public learns about exoplanets through a variety of sources. How has the input from exoplanet science influenced the public's perceptions of Earth's uniqueness in the universe and shaped views on the potential universality of life?

From a historic perspective, our view about 'other worlds' away from the Sun or about life in the Universe has changed from one of complete speculation to a semi-empirical one. As the best-known tool for this attempted quantification remains Drake's famous equation (Drake 1965, 2011). This equation, whose original form estimates the abundance of detectable intelligent life, is based on the multiplication of several factors, of which only one, the rate of star formation in our Galaxy, could be estimated reasonably well when the equation was originally presented. Other factors, such as the fraction of planets with life that develop technological civilizations, or the length of time over which such civilizations release detectable signals, remain to date entirely speculative. Our increasing knowledge on exoplanets has however raised two of the equation's factors, namely the fraction of stars with planets, and the number of habitable planets around such stars, to estimates that may be reliable to a factor of 'a few'. These estimates constitute one of the major advances of the past 20 years of research on exoplanets, leading for example to an estimate (Petigura et al. 2013) that about 6% of Sun-like stars have Earth-like planets .

Currently we are able to label planets as 'Earth-like' solely based on their principal physical parameters without crucial information for habitability, such as the presence of water or the type of atmosphere. The quoted 6% is therefore an upper limit for habitable Earth-like planets. The habitability of planets that are not Earth-like is hypothetical but in many cases reasonable (e.g. on planets around low-mass stars or on planet-sized moons); such objects might even provide the vast majority of habitats in the Universe (see also chapters "Habitability of Planets in Binary Star Systems", and "Habitability in Brown Dwarf Systems"). In either case is seems safe to assume that habitable planets (in the sense of fulfilling all requirements to develop life) are frequent, and that life – evolved or not – will be frequent as well, unless the unknown factors of Drake's equation are minuscule small. In that sense, Frank & Sullivan (2016) estimate that there is at least one other technological species in the observable Universe, unless the probability that a habitable zone planet develops a technological species is below $10^{-24}$ (see also chapter Where Life May Arise: Habitability).

These more general results from exoplanet science, together with the numerous individual findings of potentially habitable planets have led to the perception among the informed public that life in the Universe might be frequent and that its discovery is a question of "when" rather than "if". For instance, a survey in 2020 found that nearly 65% of Americans believe there is intelligent life on other planets (Kennedy & Lau 2021).
At the same time, activities for more direct detections of extraterrestrial life have come within technological reach. Not only classical SETI searches for technological signatures, but also searches for biosignatures on planets but within and outside of our Solar System. Such investigations are supported by a wide interest among the public, which is also essential for the support of the many ambitious projects in the coming decades. These projects in turn will advance the entire field of exoplanet science and humanity's understanding of the origin and presence of life in the Universe. And lastly, the public interest and fascination with current findings from exoplanets science may better prepare society for the day when we find out that we are not alone.

## Acknowledgements

The author acknowledges support from the Spanish Research Agency of the Ministry of Science and Innovation (AEI-MICINN) under grant PID2019-107061GB-C66, DOI: 10.13039/501100011033. This publication has made use of the Astrophysics Data System, funded by NASA under Cooperative Agreement 80NSSC21M00561.

## References

Boyajian TS (2016) Planet Hunters IX. KIC 8462852 - where's the flux? MNRAS 457, 3988-4004

Charbonneau D et al. (2000) Detection of Planetary Transits Across a Sun-Like Star. ApJ 529, L45

Christiansen JL et al. (2018) The K2-138 System: A Near-resonant Chain of Five Sub-Neptune Planets Discovered by Citizen Scientists. AJ 155, 57

Collins KA, Kielkopf JF, Stassun KG, Hessman FV (2017). AstroImageJ: Image Processing and Photometric Extraction for Ultra-precise Astronomical Light Curves. AJ 153, 77

Deeg HJ et al. (2009) Ground-based photometry of space-based transit detections: photometric follow-up of the CoRoT mission. A&A 506, 343

Deeg HJ, Belmonte JA (2018) Impact of Exoplanet Science in the Early Twenty-First Century. Handbook of Exoplanets (1st edition), ISBN 978-3-319-55332-0

Deeg HJ et al. (2020) Maintaining the Ephemeris of 20 CoRoT Planets: Transit Minimum Times and Potential Transit Timing Variations. Journal of the American Association of Variable Star Observers 48, 201

Deeg HJ, Alonso R (2024) Ground-based photometric follow-up for exoplanet detections with the PLATO mission. Contributions of the Astronomical Observatory Skalnaté Pleso 54, 142

Dragomir D et al. (2020) Securing the Legacy of TESS through the Care and Maintenance of TESS Planet Ephemerides. AJ 159:219

Dennis, A (2016), Exoplanet Observing by Amateur Astronomers, http://astrodennis.com/

Drake F (1965) The radio search for intelligent extraterrestrial life. In: Mamikunian, G., Briggs, M.H. (Eds.), Current Aspects of Exobiology. Pergamon, New York, pp.323-345

Drake F (2011) The search for extra-terrestrial intelligence. Philosophical Transactions of the Royal Society of London A: Mathematical, Physical and Engineering Sciences 369, 633-643.

ESA (2024) ESA science missions. https://www.cosmos.esa.int/our-missions , accessed on 2024 Sep 12

Frank A, Sullivan WT (2016), A New Empirical Constraint on the Prevalence of Technological Species in the Universe. Astrobiology, 16:359-362

Gary B (2010) Exoplanet Observing for Amateurs, 2nd edition, Reductionist Publications, available at http://brucegary.net/book_EOA/ExoplanetObservingAmateurs2ndEdition.zip

Ge J et al. (2022) ET White Paper: To Find the First Earth 2.0. eprint arXiv:2206.06693

Ge J et al. (2024) Search for a Second Earth – the Earth 2.0 (ET) Space Mission. Chinese Journal of Space Science, Volume 44, Issue 3: 400 – 424. DOI 10.11728/cjss2024.03.yg05

Kennedy C, Lau A (2021) Most Americans believe in intelligent life beyond Earth; few see UFOs as a major national security threat. Pew Research Center, https://pewrsr.ch/3w6Bgcx

Kissler-Patig M, Lyubenova M (eds) 2011. An Expanded View of the Universe: Science with the European Extremely Large Telescope. https://www.eso.org/sci/facilities/eelt/science/doc/eelt_sciencecase.pdf

Klagyivik P et al (2021) Orbital period refinement of CoRoT planets with TESS observations. Frontiers in Astronomy and Space Sciences 8, id. 210

Kokori A et al. (2022) ExoClock project: an open platform for monitoring the ephemerides of Ariel targets with contributions from the public. Exp Astron 53, 547

Kokori A et al. (2023) ExoClock Project. III. 450 New Exoplanet Ephemerides from Ground and Space Observations. ApJS 265, id. 4

Lintott C (2019). The Crowd and the Cosmos: Adventures in the Zooniverse. Oxford, New York: Oxford University Press. ISBN 9780198842224

Lund MB, Pepper J, Stassun K (2015) Transiting Planets With LSST. I. Potential for LSST Exoplanet Detection. AJ 149, id. 16

Marchis F et al. (2020) Unistellar eVscopes: Smart, portable, and easy-to-use telescopes for exploration, interactive learning, and citizen astronomy. Acta Astronautica 166, 23

McCullough PR et al. (2007) A Transiting Planet of a Sun-like Star, ApJ 648, 1228-1238

Motl D (2024) C-Munipack version 2.1.37. https://c-munipack.sourceforge.net/

NASA (2024) Flight Programs. https://science.nasa.gov/astrophysics/programs/flight-programs/ , accessed on 2024 Sep 12

Noguer FR et al. (2024) Enhancing Exoplanet Ephemerides by Leveraging Professional and Citizen Science Data: A Test Case with WASP-77 A b. PASP 136:064401

Onuchukwu CC et al. (2024) Introducing astronomy clubs in secondary schools - an effective way to boost interest in STEM education. Journal of Educational Research & Practice, Vol. 4 No. 8

Peluso DO et al. (2023) The Unistellar Exoplanet Campaign: Citizen Science Results and Inherent Education Opportunities. PASP 135:015001

Petigura EA, Howard AW, Marcy GW (2013) Prevalence of Earth-size planets orbiting Sun-like stars. Proceedings of the National Academy of Sciences, vol. 110, issue 48, pp. 19273

Poddany S, Brat L, Pejcha O (2010) Exoplanet Transit Database. Reduction and processing of the photometric data of exoplanet transits. New Astronomy 15, 297-301

Rauer H et al. (2014) The PLATO 2.0 mission. Experimental Astronomy 38:249

Rauer H et al (2024) The PLATO Mission. eprint arXiv:2406.05447

Schuetz M (2018) Recent Developments at the Boquete Optical SETI Observatory and Owl Observatory. eprint arXiv:1809.01956

Schwamb ME et al. (2012) Planet Hunters: Assessing the Kepler Inventory of Short-period Planets. ApJ 754: 129

Schwamb ME et al. (2013) Planet Hunters: A Transiting Circumbinary Planet in a Quadruple Star System, ApJ, 768, 127

Sgro LA et al. (2024) Confirmation and Characterization of the Eccentric, Warm Jupiter TIC 393818343 b with a Network of Citizen Scientists. AJ 168, id. 26

Skidmore W (editor) (2015) Thirty Meter Telescope: Detailed Science Case: 2015, http://www.tmt.org, document TMT.PSC.TEC.07.007.REL02.

Tamburo P, Muirhead PS, Dressing CD (2023) Predicting the Yield of Small Transiting Exoplanets around Mid-M and Ultracool Dwarfs in the Nancy Grace Roman Space Telescope Galactic Bulge Time Domain Survey. AJ 165, id. 251

Team-FreeAstro (2024) Siril - image processing tool for astronomy and other. https://siril.org/ (Acessed 27 Sep 2024)

Wright JT, Sigurðsson S (2016) Families of Plausible Solutions to the Puzzle of Boyajian's Star. ApJ 829, 1L

Yüzgeç S, Okuşluk F (2023) The Impact of STEM-Based Astronomy Activities on Secondary School Students' Attitudes towards STEM and Astronomy. Journal of Education, Theory and Practical Research, Vol 9, 1. DOI:10.38089/ekuad.2023.129

Zellem RT et al (2019) Engaging Citizen Scientists to Keep Transit Times Fresh and Ensure the Efficient Use of Transiting Exoplanet Characterization Missions. Astro2020: Decadal Survey on Astronomy and Astrophysics, science white papers, no. 416; Bulletin of the American Astronomical Society, Vol. 51, Issue 3, id. 416.